\documentstyle[preprint,prl,
aps,epsf]{revtex}

\begin{document}

\draft

\title{Proton Spin Relaxation Induced by Quantum Tunneling in Fe8 Molecular 
Nanomagnet}

\author{Miki Ueda, Satoru Maegawa
}
\address{
Graduate School of Human and Environmental Studies, Kyoto University, 
Kyoto 606-8501, Japan\\
}
\author{Susumu Kitagawa}
\address{
Faculty of Engineering, Kyoto University, Kyoto 606-8501, Japan
}
\maketitle

\begin{abstract}

The spin-lattice relaxation rate $T_{1}^{-1}$ and NMR spectra of $^1$H
in single crystal molecular magnets of Fe8
have been measured down to 15 mK.
The relaxation rate $T_1^{-1}$ shows a strong temperature dependence down to 400 mK.  
The relaxation is well explained in terms of the  
thermal transition of the iron state between the discreet energy levels 
of the total spin $S=10$.
The relaxation time $T_1$ 
becomes temperature independent below 300 mK and is longer than 100 s.
In this temperature region stepwise recovery of the $^1$H-NMR signal after 
saturation was observed depending on the return field of the sweep field. 
This phenomenon is attributed to resonant quantum 
tunneling at the fields where levels cross and is discussed in terms of 
the Landau-Zener transition.

\end{abstract}


Recently, molecular nanomagnets have attracted much attention to study
quantum mechanical phenomena in the macroscopic system owing to their 
identical size, the well-defined structure 
and a well-characterized energy structure.~\cite{rf:1,rf:2,rf:3}
The molecular magnet 
[(C$_{6}$H$_{15}$N$_{3}$)$_{6}$Fe$_8$O$_2$(OH)$_{12}$]Br$_{7}$(H$_{2}$O)Br$\cdot$8H$_{2}$O, 
abbreviated Fe8, is a representative compound in which quantum tunneling 
of magnetization (QTM) has been observed
as the temperature-independent recovery of magnetization below 400 
mK.~\cite{rf:4,rf:5,rf:7,rf:8}

The molecular magnet Fe8 consists of eight Fe$^{3+}$ ions with spin 
$s=5/2$ in each molecule.  The magnetic interactions between the spins 
in the molecule are antiferromagnetic and their magnitudes are $22 
\sim 170$ K,~\cite{rf:10} while the magnetic interactions between the 
molecules are negligibly small.  The magnetic properties of this 
compound at low temperatures have been described by a total 
spin of $S=10$ for each molecule, in which six spins are parallel to 
each other and the remaining two spins are antiparallel to the six 
spins.~\cite{rf:10,rf:17} The spin Hamiltonian in the 
field \mbox{\boldmath $H$} is expressed by
\begin{eqnarray}
    {\cal H}=D S_z^2+E(S_x^2-S_y^2)+g \mu_{\rm B} \mbox{\boldmath $S$} \cdot 
    \mbox{\boldmath $H$},
    \label{eq:1}
\end{eqnarray}
where $D$ and $E$ are the easy axis and the in-plane anisotropy, 
respectively.~\cite{rf:4} We have determined the values of $D=-0.276$ 
K and $E=-0.035$ K by magnetization measurement
on a single crystal.~\cite{rf:17}
When there is no magnetic field, the anisotropy stabilizes
the degenerate spin states of $m=\pm 10$ at low temperatures.
These states correspond to opposite directions of the magnetization
in the classical sense. 
The energy barrier caused by the anisotropy for reversal of 
the magnetization between $m=\pm 10$ is reported to be 25 K.~\cite{rf:4}

Quantum tunneling of magnetization in Fe8 has been observed only 
by magnetization measurements.~\cite{rf:4,rf:5,rf:7,rf:8} 
In order to study the spin dynamics in Fe8 from a microscopic 
point of view we performed NMR experiments and found a stepwise 
recovery of the $^1$H spin echo signal due to resonant quantum 
tunneling at the level crossing fields.

We synthesized single crystals of Fe8, following the 
method reported previously.~\cite{rf:9} The sizes of the  
crystals used for the experiments were about 4 $\times$ 2 $\times$ 1 
mm$^3$. 
The NMR spectrum and the spin-lattice relaxation rate $T_1^{-1}$ of 
$^1$H in Fe8 were measured by the coherent pulsed NMR 
method in external magnetic fields up to 5.4 T. The 
spectra were obtained by measuring the spin-echo intensity with
sweeping the field at a fixed frequency.
The relaxation rate was obtained by measuring the recovery of 
the spin echo intensity as a function of the time after
saturation of the $^1$H spins using comb pulses.
The experimental temperature was lowered down to 15 mK using a 
$^3$He-$^4$He dilution refrigerator.  The samples were sealed with 
$^4$He gas in a cell made of PCTFE plastic(polychlorotrifluoroethylene).
The cell was set in the mixing chamber of the dilution refrigerator.

The NMR spectrum at room temperature is narrow with the width smaller 
than 50 Oe, while the spectrum becomes broader and shows structure at 
lower temperatures. The structure is caused by freezing of the 
iron magnetization and the existence of many $^1$H sites.

\begin{figure}
\centerline{
\epsfxsize=12cm
\epsfbox{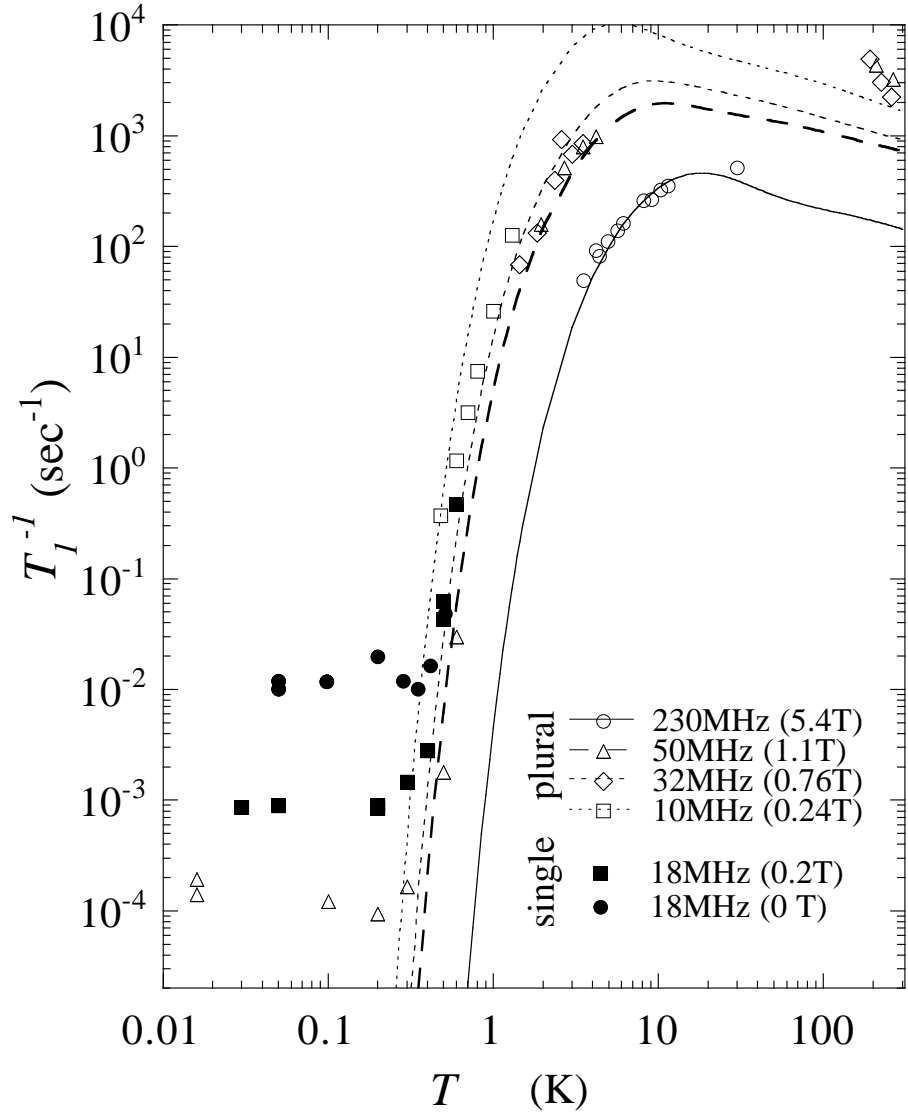}}                                                         
\caption{Temperature
dependence of  
relaxation rate $T_{1}^{-1}$ for the samples composed of plural single 
crystals and for a single crystal. 
Lines denote calculated values.
}
\label{fig:1}
\end{figure}

Figure~\ref{fig:1} shows the temperature  
dependence of the relaxation rate $T_1^{-1}$.
The rate $T_{1}^{-1}$ decreases steeply over six decades with lowering the 
temperature from 10 K to 400 mK, and decreases with increasing field.
Below 300 mK $T_{1}^{-1}$ is almost temperature 
independent and has a strong site dependence.

\begin{figure}
\centerline{
\epsfxsize=14cm
\epsfbox{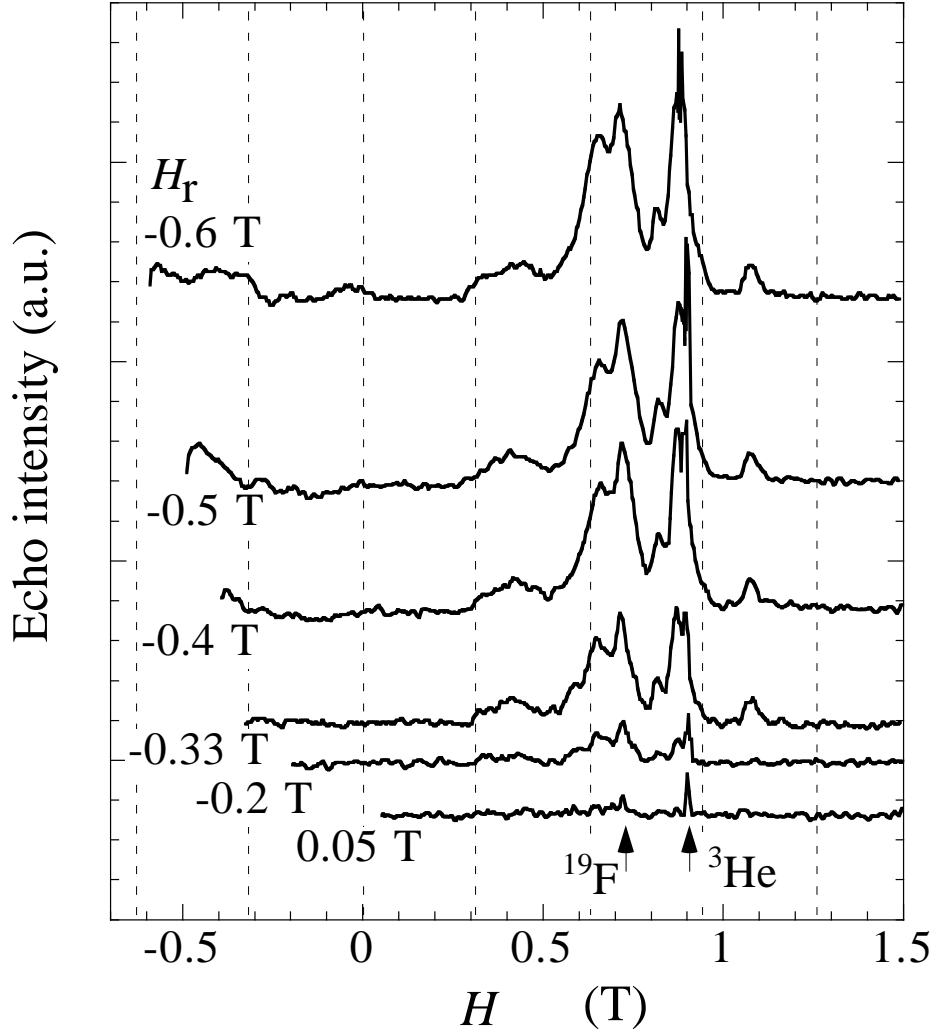}}                                                         
\caption{
NMR spectra of single crystal Fe8, which were taken with 
increasing the field from $H_{r}$ after the saturation.
$T=150$ mK, $f=29$ MHz, $dH/dt=0.9$ T/min and $\theta=50^\circ$.
Broken lines show the level crossing fields. 
}
\label{fig:2}
\end{figure}

Figure~\ref{fig:2} shows NMR spectra of the single crystal at 150 mK.
The field was applied so as to make an angle $\theta$ of $50^{\circ}$ 
from the easy axis and to be within the $ab$-plane. 
In this experiment we first saturated the $^1$H spin system by the 
comb pulses at a fixed frequency, sweeping the field up and down
between 0.1 T and 1.5 T for more than three times, until there was no 
observation of the echo signals.
Then the field was 
decreased to a certain field $H_r$ with a constant sweep rate 
$dH/dt$.  Immediately after arriving at $H_r$, the field was increased 
with the same sweep rate and the spectrum was taken at a fixed 
repetition time.
No signals were expected to be observed, because the returning duration 
after the saturation is planned to be enough shorter than the 
relaxation time that is longer than 100 s.  Indeed no signals except 
from $^{19}$F in the material of the sample cell and $^3$He 
in the mixing chamber were observed when the field was returned at 0.05 
T, as shown in the lowest spectrum in Fig.~\ref{fig:2}. However, 
signals were observed when the return fields were negative.  
The intensities of the signals were increased with decreasing
return field.

\begin{figure}
\centerline{
\epsfxsize=12cm
\epsfbox{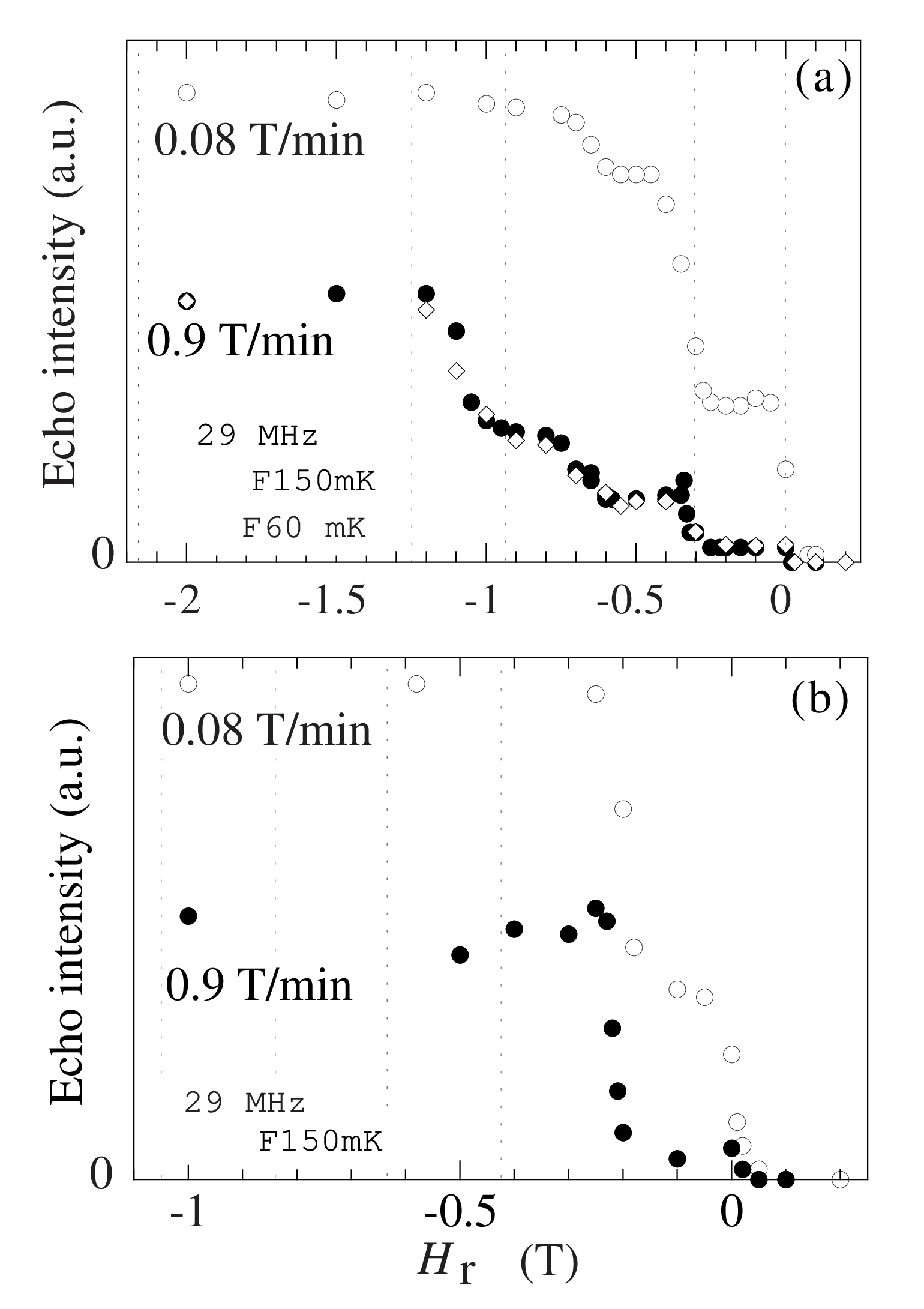}}                                                         
\caption{
Return field $H_{r}$ dependence of echo intensity at 
field of (a) 0.45 T, $\theta=50^\circ$ and (b) 0.60 T, $\theta=0^\circ$.
Broken lines show calculated level crossing fields.
The echo intensity at 60 mK in (a) is normalized with that for 150 mK at 
$H_{r}=-2.0$ T.
}
\label{fig:3}
\end{figure}

\begin{figure}
\centerline{
\epsfxsize=12cm
\epsfbox{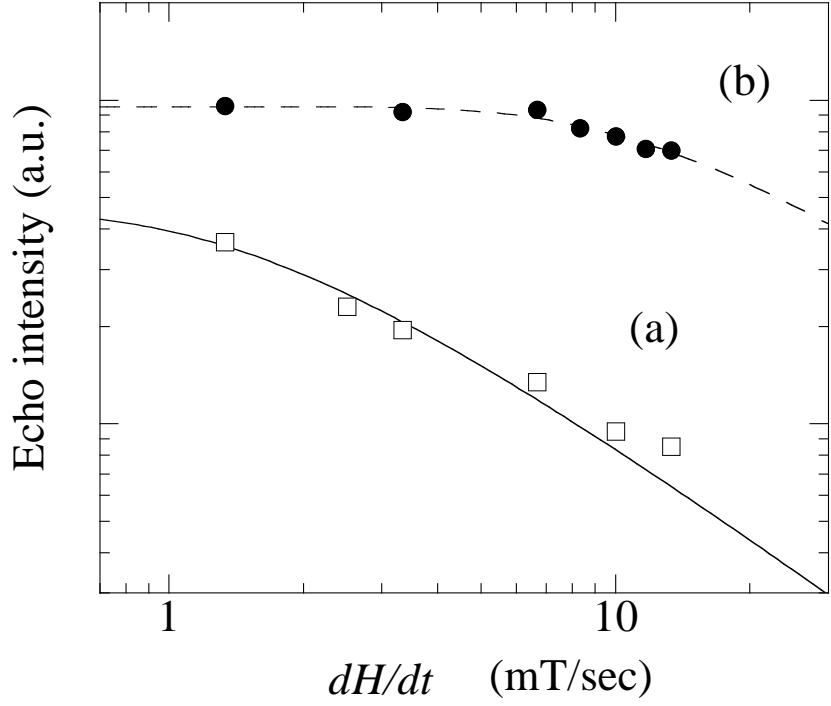}}                                                         
\caption{
Sweep rate $dH/dt$ dependence of echo intensity 
at 0.60 T when (a) $H_{r}=-0.1$ T and (b) $H_{r}=-0.3$ T.
The field is applied parallel to the easy axis.
The solid and broken lines show calculated values by eq.(\ref{eq:8}) and 
eq.(\ref{eq:9}), respectively.
}
\label{fig:4}
\end{figure}

Figures~\ref{fig:3}(a) and (b) show the return field dependence of 
the echo intensity.  Figure~\ref{fig:3}(a) was obtained in the case 
when the field was applied  
with $\theta=50^{\circ}$ 
within the $ab$-plane.  
Figure~\ref{fig:3}(b) is the case when the field was parallel to the 
easy axis $(\theta
=0^{\circ})$.
The echo intensities were picked up at fields of 0.45 T 
and 0.60 T for Fig.~\ref{fig:3}(a) and (b), respectively. The 
intensity increases sharply with steps at 0, $-0.31$,$-0.63$ and 
$-1.00$ T for Fig.~\ref{fig:3}(a), while for Fig.~\ref{fig:3}(b)
the steps are at +0.02 and $-0.20$ T. The increase in the 
intensity with the steps cannot be explained by recovery due to the 
spin-lattice relaxation time shown in Fig.~\ref{fig:1}.  Moreover the 
intensity depends on the sweep rate of the field.
Figure~\ref{fig:4} shows the sweep rate dependence of the 
intensity at a field of 0.60 T when the field was applied parallel 
to the easy axis and the return fields were $-0.1$ T and $-0.3$ T.  
The intensity is large when the sweep rate is slow.

First we analyze the relaxation rate.
The temperature dependent relaxation rates of a molecular magnet 
Mn12 were first discussed by Lascialfari {\it et al.}~\cite{rf:11}
In general $T_{1}^{-1}$ is given by the Fourier transform of the 
correlation function for fluctuating transverse local fields 
$h_{\pm}(t)$ at nuclear sites, and is expressed as
\begin{eqnarray}
\frac{1}{T_{1}} & = & \frac{\gamma_{N}^{2}}{2} \int \left<h_{\pm}(t) 
	h_{\mp}(0)\right>e^{i\omega_{L} t} dt ,
\label{eq:10}
\end{eqnarray}
where $\gamma_{N}$ is the nuclear gyromagnetic ratio and $\omega_{L}$ 
is the Larmor frequency.  When the correlation is assumed to 
be described by an exponential function with a life time $\tau_{m}$ 
in the $m$ state, the correlation
can be expressed as
\begin{eqnarray}
    \left<h_{\pm}(t)h_{\mp}(0)\right>=
	\sum_{m=-10}^{+10}{\left<\Delta h_{\pm}^2\right>e^{-t/\tau_{m}} \frac{e^{-E_{m}/k_{\rm B} T}}{Z}},
\label{eq:2}
\end{eqnarray}
where $E_{m}$ is the energy of the eigenstate $m$
and $Z$ is the partition function.
The lifetime $\tau_{m}$ for the $m$ state is governed by the 
spin-phonon interaction ${\cal H}_{sp}$ and is expressed by the transition 
probabilities $p_{m \rightarrow m-1}$ for the transition from the $m$ to 
the $m-1$ state and $p_{m \rightarrow m+1}$ from the $m$ to the $m+1$ 
state,
as follows~\cite{rf:12}
\begin{eqnarray}
& &\frac{1}{\tau_{m}}=p_{m \rightarrow m-1}+p_{m \rightarrow m+1}  \nonumber \\
 & &= \cases{
\displaystyle\frac{C\Delta E_{m}^3}{e^{\Delta E_{m}/k_{\rm B} T}-1} +
\displaystyle\frac{C\Delta E_{m+1}^3}{1-e^{-\Delta E_{m+1}/k_{\rm B} T}} ,
		{\rm{\;(for\;}}m>0) \cr 
\displaystyle\frac{C\Delta E_{m+1}^3}{e^{\Delta E_{m+1}/k_{\rm B} T}-1} +
\displaystyle\frac{C\Delta E_{m}^3}{1-e^{-\Delta E_{m}/k_{\rm B} T}} ,
		{\rm{\;(for\;}}m<0) \cr} \label{eq:4}
\end{eqnarray}
where $\Delta E_{m}=|E_{m-1}-E_{m}|$.
The parameter $C$ is given by
\begin{eqnarray}
	C=\frac{3}{2 \pi \rho v^5 \hbar^4} |\left<m|{\cal H}_{sp}|m \pm 
	1\right>|^2,
\label{eq:5}
\end{eqnarray}
where 
$v$ is the phonon velocity and $\rho$ is the specific mass.
Thus the spin-lattice relaxation rate is obtained as
\begin{eqnarray}
	\frac{1}{T_{1}} & = & \frac{A}{Z} \sum_{m=-10}^{+10}
	{\frac{\tau_{m}e^{-E_{m}/k_{\rm{B}} T}}{1+\omega_{L}^{2} \tau_{m}^{2}}} , 
\label{eq:6}
\end{eqnarray}
where $A=\gamma_{N}^{2} <\Delta h_{\pm}^{2}>$. 

The relaxation rates calculated by using eq.(\ref{eq:6}) with
fitting parameters $A=4 \times 10^{12}$ rad/s$^2$ and $C=5 \times 
10^5$ Hz/K$^3$ are shown for several fields in 
Fig.~\ref{fig:1}. The experimental results of the temperature and 
field dependence 
above 400 mK are well 
reproduced over six decades by this calculation.  This means that 
relaxation above 400 mK is dominated by  
thermal fluctuations resulting from the transitions of iron spins 
between neighboring states 
due to spin-phonon interactions.  The deviations at high temperatures must 
be caused by the contribution from higher energy levels which are 
not described by the simplified spin model with $S=10$.

The observed 
$T_{1}^{-1}$ below 300 mK deviates
from the calculated values and becomes temperature independent.
This may be related to the temperature 
independent recovery of magnetization that has been observed in SQUIDs 
below 400 mK at non-level crossing fields,~\cite{rf:4} and may be
attributed to the quantum effect.

\begin{figure}
\centerline{
\epsfxsize=12cm
\epsfbox{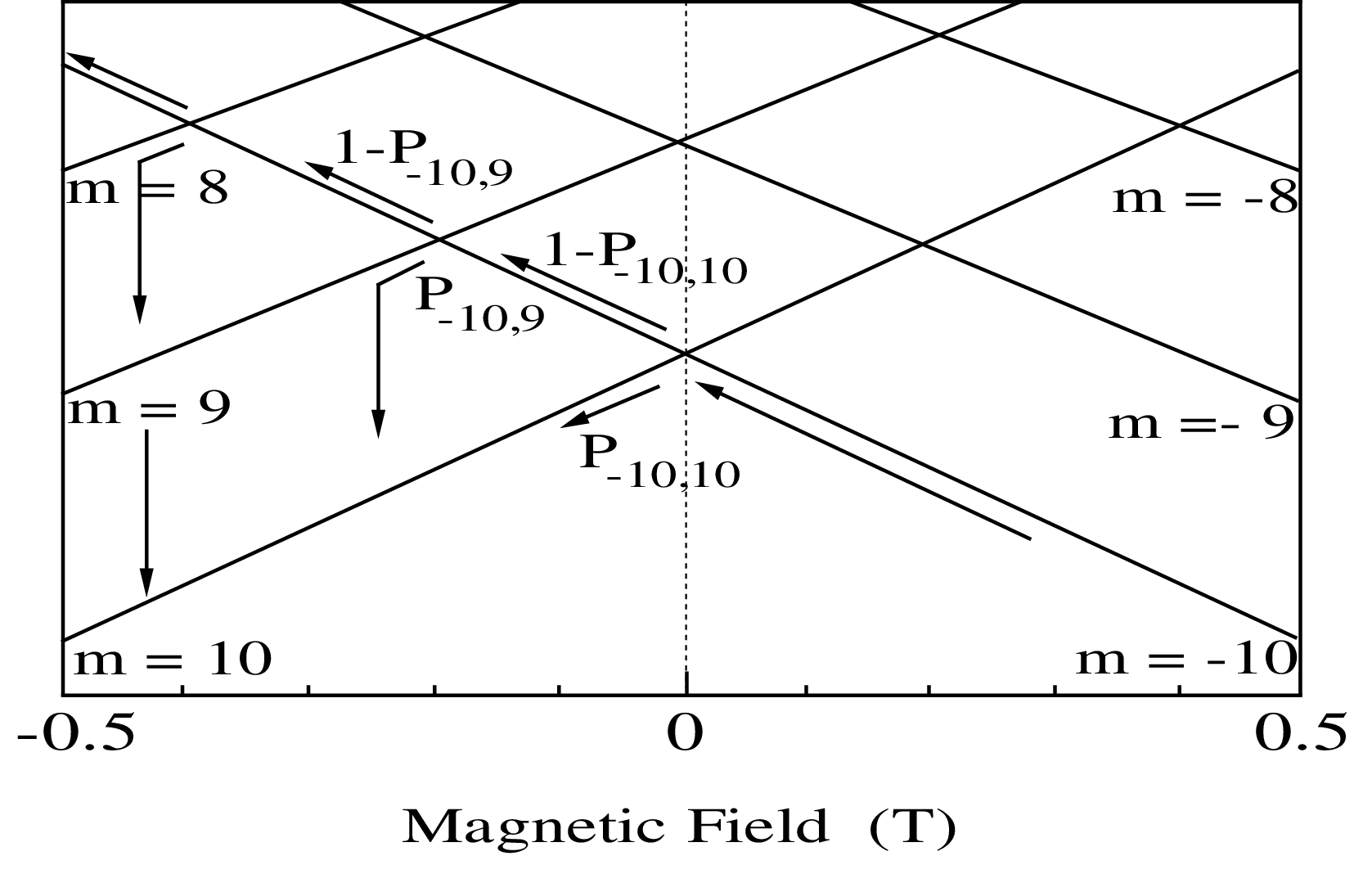}}                                                         
\caption{
Schematic energy level diagram of the spin system and the transitions.
}
\label{fig:5}
\end{figure}

Next we discuss the stepwise behavior  
of the echo intensity, shown in Fig.~\ref{fig:3}. 
Figure~\ref{fig:5} shows schematically the field dependence of the 
energy levels of the iron spin system described by eq.(\ref{eq:1}) and the 
transitions of the 
states when the field is decreased.
In the experiment, when the field is positive at low temperatures, all 
the spins are in the $m=-10$ state.
When the field is decreased to zero, the level of the $m=-10$ state crosses with
that of the $m=10$ state, and the spins in the $m=-10$ state would change to 
the $m=10$ state by quantum tunneling with a probability $P_{-10,10}$.
Further decrease of the field causes the energy levels to cross again
and the transition to occur again.
The transition at the point where the levels cross has been calculated
by Landau and Zener, and the  
probability from the $m$ state to the $m'$ state in this 
case is expressed~\cite{rf:19,rf:13,rf:14} as
\begin{eqnarray}
	P_{m,m'} & = & 1-\exp(-\frac{\pi \Delta_{m,m'}^2}{2 \hbar g \mu_{\rm{B}}
	         |m-m'| dH/dt}) ,
\label{eq:7}
\end{eqnarray}
where $\Delta_{m,m'}$ is the tunneling gap at the 
level crossing.
These transitions 
would induce fluctuations of the local field at 
proton sites and cause extra relaxation of the nuclear spins.

Immediately after quantum tunneling, the spin state goes to the 
lower energy levels, following the Boltzman distribution.
The life time $\tau_m$  
is estimated to be less than $10^{-7}$ sec from the 
measured relaxation time and eq.(\ref{eq:6}).
It should be noted that these thermal transitions can occur only 
between  
states with the same sign of $m$, while quantum tunneling occurs 
between states with opposite sign of $m$.

The measured period of the stepwise behavior for $\theta=50^\circ$
was 0.32 T, 
while that for $\theta=0^\circ$ was 0.22 T.
The period of level crossing fields is expressed approximately 
as $\Delta H=D/g\mu_{\rm{B}}\rm{cos}\theta$ and 
the values are  
0.31 T and 0.21 T for $\theta=50^\circ$ and $0^\circ$, respectively.
The values coincide fairly well with the experimental results.
These results clearly indicate 
that the sudden recovery of the $^1$H spins is caused by
resonant quantum tunneling of the iron magnetization at the 
level crossing fields.

When the iron spin state in a certain molecule changes through
tunneling, the $^{1}$H spins in the 
molecule would be relaxed immediately. 
The iron spins which have tunneled in zero or negative fields arrive 
at the $m=10$ state and have possibility to tunnel again in the zero or
positive fields. 
If it is assumed that the iron spins which have experienced tunneling 
once at least contribute the proton relaxation, the echo intensities 
$I_i$ at the monitoring field after the field returning would be
expressed as 
\begin{eqnarray}
	I_{1} & = & I_{\rm{01}}\{1-\left(1-P_{-10,10}\right)^2\},
\label{eq:8}
\end{eqnarray}
for $H_{-10,9}<H_r<0$, and 
\begin{eqnarray}	
	I_{2} & = & 
	I_{\rm{02}}\{1-\left(1-P_{-10,10}\right)^2\left(1-P_{-10,9}\right)^2\},
\label{eq:9}
\end{eqnarray}
for $H_{-10,8}<H_r<H_{-10,9}$.
The measured sweep rate dependence of echo intensities
was fitted to eqs.(\ref{eq:8}) and (\ref{eq:9}) with 
eq.(\ref{eq:7}),
$\Delta_{-10,10}=3.52\times10^{-7}$ K and
$\Delta_{-10,9}=9.66\times10^{-7}$ K, as shown in Fig.~\ref{fig:4}.
The agreement is fairly well and these 
values of $\Delta_{-10,10}$ and $\Delta_{-10,9}$ are close to 
those reported from magnetization measurements.~\cite{rf:7,rf:8} 

As is found in Fig.~\ref{fig:3}(b), the echo intensities for 
$H_{r}<-0.25$ T at a sweep rate of 0.9 T/min
remain constant and smaller than those for $dH/dt=0.08$ T/min.
The stepwise recovery was not observed for the lower crossing fields.
An intensity of the signals which were
measured at the slow sweep rate of 0.08 T/min with $H_r$ less than 
$-0.6$ T corresponds to the intensity of the signal which recovered completely. 
This intensity was obtained for the fast rate of 0.9 T/min by sweeping 
the field up and down three times
between $-0.1$ T and $-0.6$ T.
However, the intensity after sweeping up and down three times at lower 
fields between $-0.6$ T and $-1.0$ T with $dH/dt=0.9$ T/min 
remained small.
This suggests that tunneling does not occur at fields lower than
$-0.6$ T, though the reason is not clear.

In conclusion, 
relaxation rate $T_{1}^{-1}$ 
of Fe8  
above 400 mK are dominated by the thermal 
fluctuation of the iron magnetization with spin $S=10$. The transition 
of magnetization between the states splitted by $DS_{Z}^2$ are caused 
by the spin-phonon interaction, and induces the fluctuation at 
nuclear sites.
The rate becomes temperature independent below 300 mK.
In this temperature region
the stepwise recovery of the echo intensity caused by quantum 
tunneling 
at the level crossing fields was observed in the $^1$H-NMR spectrum.
This indicates that the internal field at 
proton sites fluctuates due to resonant quantum tunneling of iron spins,
which is described by the Landau-Zener transition.

We would like to thank Prof. S. Miyashita, Prof. B. Barbara and Dr. K 
Saito for valuable discussions.
We also thank Dr. H. Miyasaka and Dr. H.-C.
Chang for their helpful advice on synthesis of the samples.

\newpage

%
%
%
%

\newpage

\newpage

\newpage

\newpage

\newpage

\end{document}